\documentclass[12pt]{article}
\usepackage{amsmath,amssymb,amsfonts,epsfig,graphicx,euscript}
\newcommand{\be}{\begin{equation}}
\newcommand{\ee}{\end{equation}}

\newcommand{\bse}{\begin{subequations}}
\newcommand{\ese}{\end{subequations}}
\newcommand{\bea}{\begin{eqnarray}}
\newcommand{\eea}{\end{eqnarray}}
\newcommand{\ba}{\begin{array}}
\newcommand{\ea}{\end{array}}

\makeatletter \@addtoreset{equation}{section}

\makeatother

\begin{document}
\baselineskip 18pt%

\begin{titlepage}

\begin{flushright}\vspace{-2cm}
{\small
IPM/P-2009/032\\
 }\end{flushright} \vspace{5 mm}

\begin{center}
\centerline{\Large{\bf{Rotating mesons in the presence }}}%
\centerline{\Large{\bf{of higher derivative corrections from gauge-string duality}}}%
\end{center}
\vspace*{5mm}
\begin{center}
{\bf M. Ali-Akbari$^1$ and K. Bitaghsir Fadafan$^2$}%
\vspace*{0.4cm}

{\it {$^1$School of Physics, Institute for Research in Fundamental Sciences (IPM)\\
P.O.Box 19395-5531, Tehran, Iran}}  \\
{E-mails: {\tt aliakbari@theory.ipm.ac.ir}}\\%
{\it {$^2$Physics Department, Shahrood University of Technology,\\
P.O.Box 3619995161, Shahrood, Iran }}\\
{E-mails: {\tt bitaghsir@shahroodut.ac.ir}}%
\vspace*{1.5cm}
\end{center}

\begin{abstract}
We consider a rotating quark-antiquark pair in $\mathcal{N}=4$
thermal plasma. By using the AdS/CFT correspondence, the properties
of this system are investigated. In particular, we study variation
of a rotating string radius at the boundary as a function of its tip
and angular velocity. We also extend the results to higher
derivative corrections {\em{i.e}}. ${\cal{R}}^2$ and ${\cal{R}}^4$
which correspond to finite coupling corrections on the rotating
quark-antiquark system in the hot plasma. In the ${\cal{R}}^4$ case
and for fixed angular velocity the string endpoints become more and
more separated as $\lambda^{-1}$ decreases. To study ${\cal{R}}^2$
corrections, rotating quark-antiquark system in Gauss-Bonnet
background has been considered. We summarize the effects of these
corrections in the conclusion
 section.%

\end{abstract}
\end{titlepage}
%
\section{Introduction}
The theory of strong nuclear interactions, QCD, exhibits various
phenomena at low energy. These phenomena are strong coupling effects
which are not visible in perturbation theory and there are no known
quantitative methods to study them (except by lattice simulations).
A new method for studying different aspects of strongly coupled
gauge theories is the AdS/CFT correspondence
\cite{MAGO,Maldacena:1997re,Gubser:1998bc,Witten:1998qj,
Witten:1998zw} which has yielded many important insights into the
dynamics of strongly-coupled gauge theories
\cite{Peeters:2007ab,Mateos:2007ay}. Methods based on AdS/CFT relate
gravity in $AdS_5$ space to conformal field theory on the
4-dimensional boundary, CFT$_4$. It has been shown that an AdS space
with a black brane is dual to conformal field
theory at finite temperature.%

This method is particulary powerful in large $N_c$ gauge theories
which are dual to weakly coupled string theories in weakly curved
spacetime, since then many computations can be explicitly performed.
Although AdS/CFT correspondence is not directly applicable to QCD,
one expects that results obtained from closely related non-abelian
gauge theories should shed qualitative (or even quantitative)
insights into analogous questions in QCD
\cite{Edelstein:2009iv,Myers:2008fv}. This has motivated much work
devoted to study various properties of thermal SYM theories like the
hydrodynamical transport quantities \cite{Gubser:2009fc}. %

The experiments of Relativistic Heavy Ion Collisions (RHIC), in
which gold nuclei collide at 200 GeV per nucleon, have produced a
strongly coupled quark-gluon plasma (QGP) \cite{Shuryak:2004cy}. In
order to find out different aspects of QGP and reproduce
experimental results, AdS/CFT technology has been used
$e.g.$\cite{Liu:2006nn,Liu:2006ug,Herzog:2006gh,CasalderreySolana:2006rq,Gubser:2006bz,Gubser:2006qh}
.%

In this paper we study a rotating quark-antiquark pair in
$\mathcal{N}=4$ thermal plasma which can be interpreted as a meson
\cite{Erdmenger:2007cm}. Based on lattice results and experiments,
it is found that the meson shows interesting behavior as the
temperature of the plasma increases. It is known that heavy quark
bound states can survive in a QGP to temperatures higher than the
confinement/deconfinement transition \cite{de Forcrand:2000jx}. We
use the gauge-string duality \cite{Gubser:2009md} to investigate the
properties of the above system. Thermal properties of \emph{static}
quark-antiquark systems have been studied in \cite{Maldacena:1998im,
Rey} in an AdS-Schwarzschild black hole setting using the AdS/CFT
correspondence. The static quark-antiquark system is modeled as an
open string hanging from the boundary in the bulk (U-shaped string)
with its endpoints representing quark and antiquark. The motion of a
quark-antiquark pair through $\mathcal{N}=4$ plasma has been
discussed in \cite{Chernicoff:2006hi,Argyres:2006vs,Sadeghi:2008ci}.
Several studies of rotating mesons and their related properties have
been done in
\cite{Kruczenski:2003be,Kruczenski:2004me,Peeters,Antipin} by means
of the AdS/CFT correspondence. In \cite{Fadafan:2008bq} rotation of
a heavy test quark has been
considered.%

It has been shown that there are two kinds of strings, long strings
and short strings, an analysis of which has shown the stability of
the latter kind
\cite{Friess:2006rk,Avramis:2006nv,Avramis:2007mv,Sfetsos:2008yr}.
We show that there are short and long strings in the case of
rotating mesons and, based on the stability analysis of binding
quark-antiquark pairs, one can show
that short strings are more energetically favorable.%

We also study the behavior of the radius $d$ of a rotating string at
the boundary by plotting $d$ first as a function of the distance
between the tip of the U-shaped string and the horizon and then as a
function of the string angular velocity $\omega$, the plots of which
may be seen in Fig. \ref{d-w1}. By a fitting approach, we give a
formula for $d$ in terms of $\omega$ in \eqref{d-w12}, which
describes only physical solutions, {\em{i.e.}} short strings. Notice
that this result is based on the
relation between the tip of the U-shaped string and $\omega$ in Fig. \ref{zminw}.%

We extend the results to higher derivative corrections which on the
gravity side correspond to finite coupling corrections on the gauge
theory side. The main motivation to consider corrections comes from
the fact that string theory contains higher derivative corrections
arising from stringy effects. In the case of $\mathcal{N}=4$ super
Yang-Mills (SYM) theory, the gravity dual corresponds to type
$\amalg$B string theory on $AdS_5\times S^5$ background. The leading
order correction in $1/\lambda$ arises from stringy correction to
the low energy effective action of type $\amalg$b supergravity,
$\alpha'^3 {\cal{R}}^4$. On the gauge theory side, computations are
exactly valid when the 't Hooft coupling constant goes to infinity
($\lambda=g_{YM}^2N\rightarrow\infty$). An understanding of how
these computations are affected by finite $\lambda$ corrections may
be essential for
more precise theoretical predictions.%

We study ${\cal{R}}^4$ and ${\cal{R}}^2$ corrections to the
properties of rotating strings by analyzing their shape for a given
angular velocity. We find that as $\lambda^{-1}$ decreases the
string endpoints become more and more separated. We also noticed
that the radius of a rotating quark-antiquark system is smaller when
${\cal{R}}^4$ correction is considered. To study ${\cal{R}}^2$
correction, we consider Gauss-Bonnet (GB) background which is the
most general theory of gravity with quadratic powers of curvature in
five dimensions. We see that the longer string has more energy than
the shorter one and by analyzing Fig. \ref{R41} and Fig. \ref{GB1},
we find that the cases with bigger 't Hooft coupling have longer
length and therefore
more energy. We summarize the effects of these corrections in the last section. %

The article is organized as follows. In the next section, we study
some properties of rotating open strings in AdS-Schwarzschild black
hole extending our analysis to higher derivative corerections in
section 3. In the last section we draw our conclusions
and summarize our results. %

\section{Rotating quark-antiquark pair at finite temperature}
In this section we study a rotating quark-antiquark system at the
boundary as an open string in the bulk space from the gauge-string
duality point of view. According to the Maldacena's conjecture,
${\cal{N}}=4$ $U(N)$ superconformal Yang-Mills theory in $3+1$
dimensions is dual to type IIB superstring theory on $AdS_5\times
S^5$ background. The latter background can be realized as a
near-horizon geometry of extremal D3-branes in type IIB superstring
theory. It is well known that in the large $N$ and strong 't Hooft
coupling limit the gauge theory is dual to type IIb supergravity on
$AdS_5\times S^5$.

At finite temperature, the large $N$ and strong coupling limit of
$D=4$ ${\cal{N}}=4$ SYM theory is dual to the near-horizon geometry
of near-extremal D3-branes in IIB superstring theory. This geometry
is given by AdS-Schwarzschild type IIb supergravity (when the other
compact $5$ dimensional manifold is $S^5$)
\begin{eqnarray}\label{adsschmetric}
 ds^{2}=\frac{u^2}{R^2}\big(-h(u)dt^{2}+ d\rho^{2}+\rho^{2}d
 \theta^{2}+dx_{3}^{2}\big)+\frac{du^{2}}{(\frac{u}{R})^2h(u)},
\end{eqnarray}
where
\begin{eqnarray}
h\left(u\right)=1-\frac{u_{h}^{4}}{u^{4}}.
\end{eqnarray}
Here $u$ is the radial direction which is bounded from below by
$u\geq u_h$ where $u_h$ refers to the location of the horizon. In
fact it maps nonperturbative problems at strong coupling onto
calculable problems of classical gravity in a five dimensional
$AdS_5$ black hole spacetime. The temperature in the hot plasma is
equal to the Hawking temperature of the AdS black hole in the
gravity dual, namely
\begin{equation}
T=\frac{u_h}{\pi\,R^2}\label{temperature}.
\end{equation}

\subsection{Rotating quark-antiquark pair as a U-shaped string}
In the follow we introduce a rotating quark-antiquark pair from
string theory and AdS/CFT point of view. In the usual fashion the
two endpoints of an open string are seen as a quark and antiquark
pair which may be considered as a meson. The open string hanging in
the bulk space and connecting two endpoints has a characteristic
U-shaped. We name $u_*$ the tip of the U-shaped string and we let it
to define the nearest point between the string and the horizon of
the black hole ($u_*\geq u_h$). Let us emphasize that for
non-physical states we would have $u_*<u_h$ \cite{Rey}.

In order to study a rotating string, we make use of the Nambu-Goto
action in the above background given by
\begin{eqnarray}
S=-\frac{1}{2\pi\alpha'}\int d\tau d\sigma\sqrt{-{\rm{det}}g_{ab }
}.
\end{eqnarray}
The coordinates $(\sigma, \tau)$ parameterize the induced metric
$g_{ab}$ on the string world-sheet. Indices $a,b$ run over the two
dimensions of the world-sheet. Let $X^\mu(\sigma, \tau)$ be a map
from the string world-sheet into spacetime and let us define $\dot X
=\partial_\tau X$, $X' =
\partial_\sigma X$, and $V \cdot W = V^\mu W^\nu G_{\mu\nu}$ where
$G_{\mu\nu}$ is the AdS black hole metric. Indices $\mu, \nu$ run
over the five dimensions of spacetime. Then
\begin{eqnarray}
 -g=-{\rm{det}}g_{ab }=(\dot X \cdot X')^2 - (X')^2(\dot X)^2.
\end{eqnarray}
The background metric is given by \eqref{adsschmetric}. Our
four-dimensional space is along $t,\rho,\theta$ and $x_3$ where the
quark-antiquark system is rotating on the $\rho,\theta$ plane with
$x_3$ the direction perpendicular to the plane of rotation. We
choose to parameterize the two-dimensional world-sheet of the
rotating string
$X^\mu(\sigma, \tau)$ according to %
\be\label{ansatz}
 X^\mu(\sigma, \tau)=\left(t=\tau,\,\,\,\rho=\sigma,\,\,\, u=u(\rho),\,\,\,\theta=\omega t\right).
\ee %

Simply, what this means is that the radius of the rotating
quark-antiquark on the probe brane changes with the fifth direction
of the bulk space as we move further into the bulk. In arriving at
the parametrization (\ref{ansatz}), we made use of the fact that the
quark-antiquark pair is in circular motion with radius $d$ at a
constant angular velocity $\omega$. Also, we assumed that the system
retains its constant circular motion at all times. Furthermore, the
ansatz (\ref{ansatz}) does not show any dragging effects which frees
us from applying a force to maintain the rigid rotation
\cite{Peeters}.

\begin{figure}[ht]
\centerline{\includegraphics[width=3.3in]{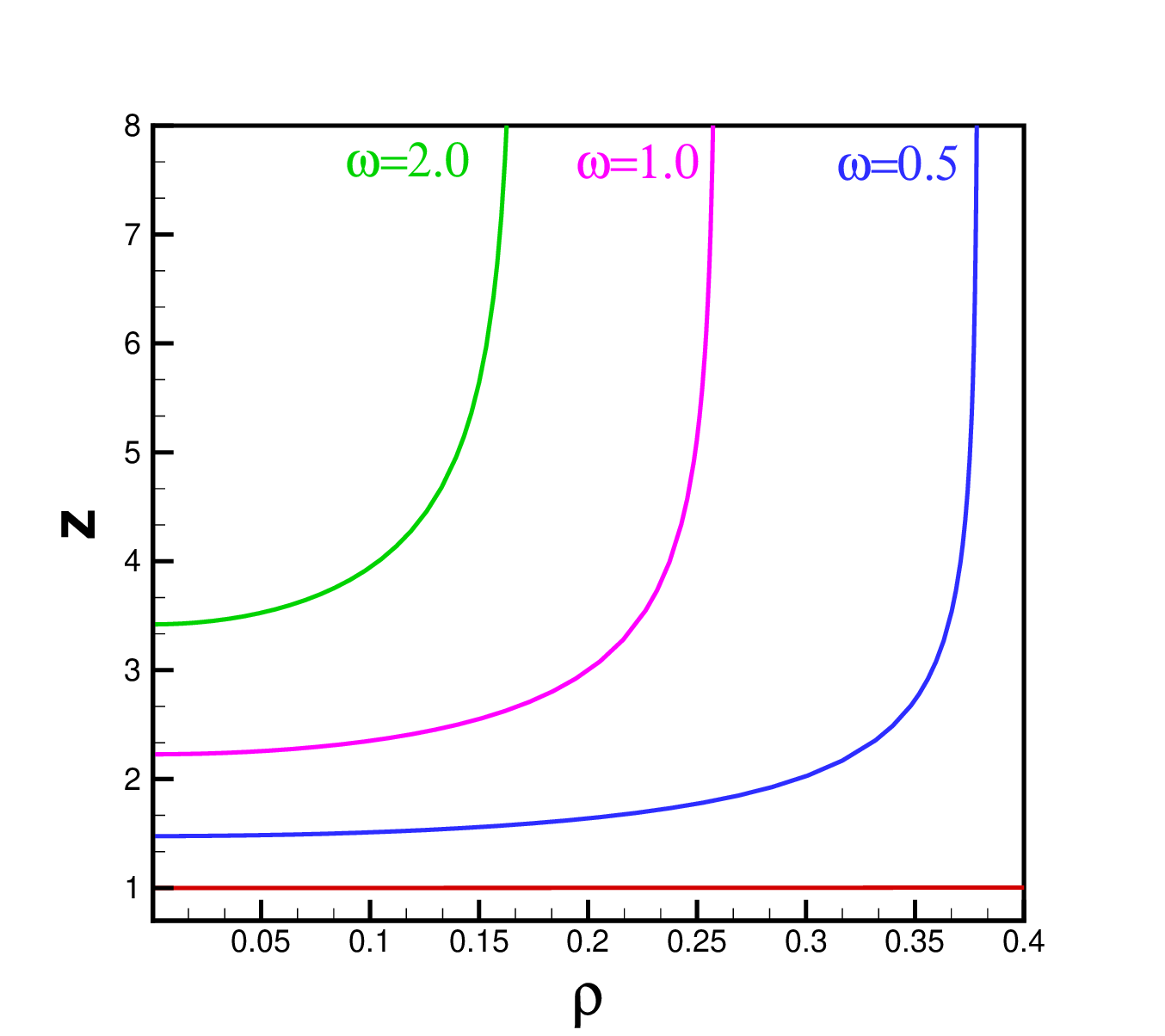}}%
\caption{Shapes of open strings for different angular velocities are
plotted. All solutions lie above the horizontal line defined by
(\ref{validity}). Notice that the positivity condition is seen as a
straight line at $z=1$.\label{z-rho} }
\end{figure}

In order to describe the rotation of a quark-antiquark pair the
end-points of the string on the probe brane must satisfy the
following boundary conditions
\begin{eqnarray}
 u(d)&=&\infty, \\
      \frac{\partial u}{\partial \rho}&=&\infty.
\end{eqnarray}
According to our ansatz (\ref{ansatz}), the Nambu-Goto action with $\alpha'=1$ becomes%
\begin{eqnarray}\label{maxaction}
 S=-\frac{1}{2\pi}\int dtd\rho\sqrt{\bigg(h(u)-\rho^2\omega^2 \bigg)
 \left(\frac{u'^2}{h(u)}+\frac{u^4}{R^4}\right)  },
\end{eqnarray}
where prime is the derivative with respect to $\rho$. We shall find
it convenient to introduce new variables%
\begin{eqnarray}\label{changeofvariable}
z=\frac{u}{u_h},\,\,\tilde{\rho}=\frac{u_h}{R^2}\rho,\,\,\tilde{\omega}=\frac{R^2}{u_h}\omega,
\end{eqnarray}
where in the new variables, prime denotes the derivative with
respect to $z$. Thus we rewrite the action \eqref{maxaction} in the
new variables as %
\begin{eqnarray}\label{action}
 S&=&-\frac{u_h}{2\pi}\int dtd\rho\sqrt{(z^4-1-\rho^2 \omega^2 z^4)
 \left(\frac{z^{\prime2}}{z^4-1}+1\right )}\nonumber\\
 &\equiv&- \frac{u_h}{2\pi} \int dtd\rho {\cal{L}}.
\end{eqnarray}
where we drop the tildes for convenience. It is evident that
positivity of the square root in \eqref{action} requires that
$z^4-1-\rho^2 \omega^2 z^4\geq 0$. This in turn means that for a
given angular velocity $\omega$,
the string solution $u(\rho)$ has to lie above the curve %
\be\label{validity}%
 u(\rho)\geq\frac{u_h}{(1-\rho^2\omega^2)^{1/4}}.
\ee %
for its action to be real. It is shown in Fig. \ref{z-rho} that all
rotating strings with different angular velocity do satisfy this
condition.

The equation of motion for $z$ follows from differentiating the
Nambu-Goto action and is given by
\begin{eqnarray}\label{max4}
 \partial_\rho \left(  \frac{z^{\prime} (1-\rho^2 \omega^2 \frac{z^4}{z^4-1})}{\sqrt{-g}} \right) -
 \frac{\partial {\cal{L}}}{\partial z}=0,\label{EOM}
\end{eqnarray}
where %
\be %
 \frac{\partial {\cal{L}}}{\partial z}=2z^3\Bigg(\frac{(z^4-1)^2+\rho^2\omega^2\big(z'^2-(z^4-1)^2\big)}
 {(z^4-1)^2\sqrt{-g}}\Bigg).
\ee %

\subsection{Numerical solutions}
The equation of motion for $z(\rho)$ in (\ref{EOM}) is nonlinear and
coupled. In the simple case of no angular velocity one can reproduce
the analysis of the \emph{static} case as given in
\cite{Maldacena:1998im,Rey}. However, for generic values of $\omega$
we cannot solve this equation analytically and we have to resort to
numerical methods.

The boundary conditions which solve \eqref{max4} are
\begin{eqnarray}\label{max5}
z(d)=20\ \ \ \  {\rm{and}}\ \ \ \ z'(d)\rightarrow \infty .
\end{eqnarray}
Physically \eqref{max5} means that string terminates orthogonally on
the brane in the boundary which in turn implies Neumann boundary
conditions. Also $z'(0)=0$ has been considered at the tip of the
string where $\rho=0$. To check the validity of our solutions, we
choose $\rho=d$ so that at $\rho=0$ we keep the condition $z'(0)=0$.

The shapes of rotating open strings, for a fixed temperature, are
shown in Fig. \ref{z-rho}. By analyzing the shapes of the string for
various angular velocities, we infer that as $\omega$ decreases the
string endpoints become more and more separated, {\em i.e.} the
radius of the open string at the boundary will increase and it will
penetrate deeper into the horizon. We would like to stress again
that rotating strings with different angular velocities have to lie
above the curve given by \eqref{validity} in order their action be
real. This is precisely what is shown in Fig. \ref{z-rho}.

\subsection{Rotating and static quark-antiquark pair }

\begin{figure}
 \centerline{ \includegraphics[width=3.2in]{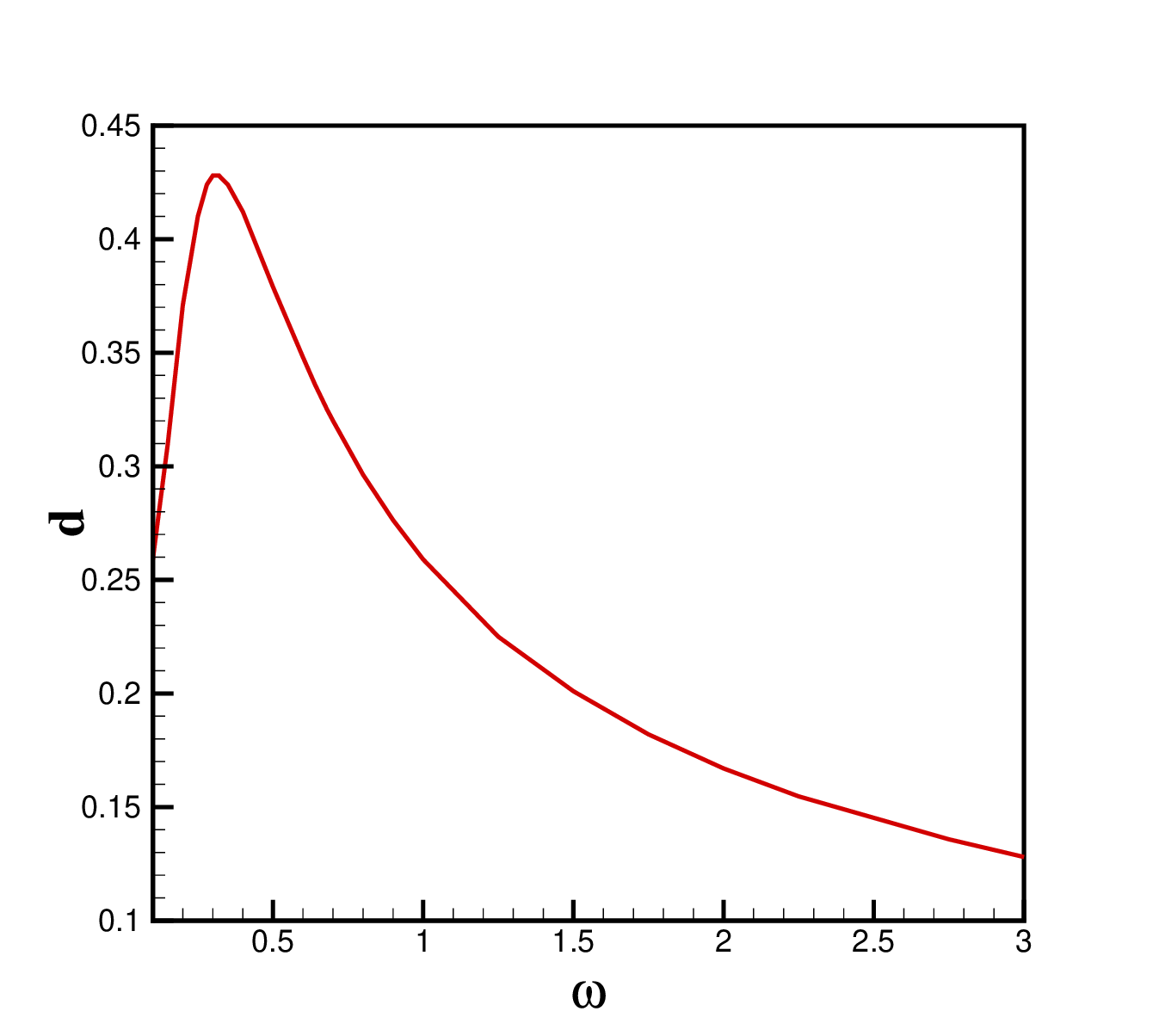} \includegraphics[width=3.2in]{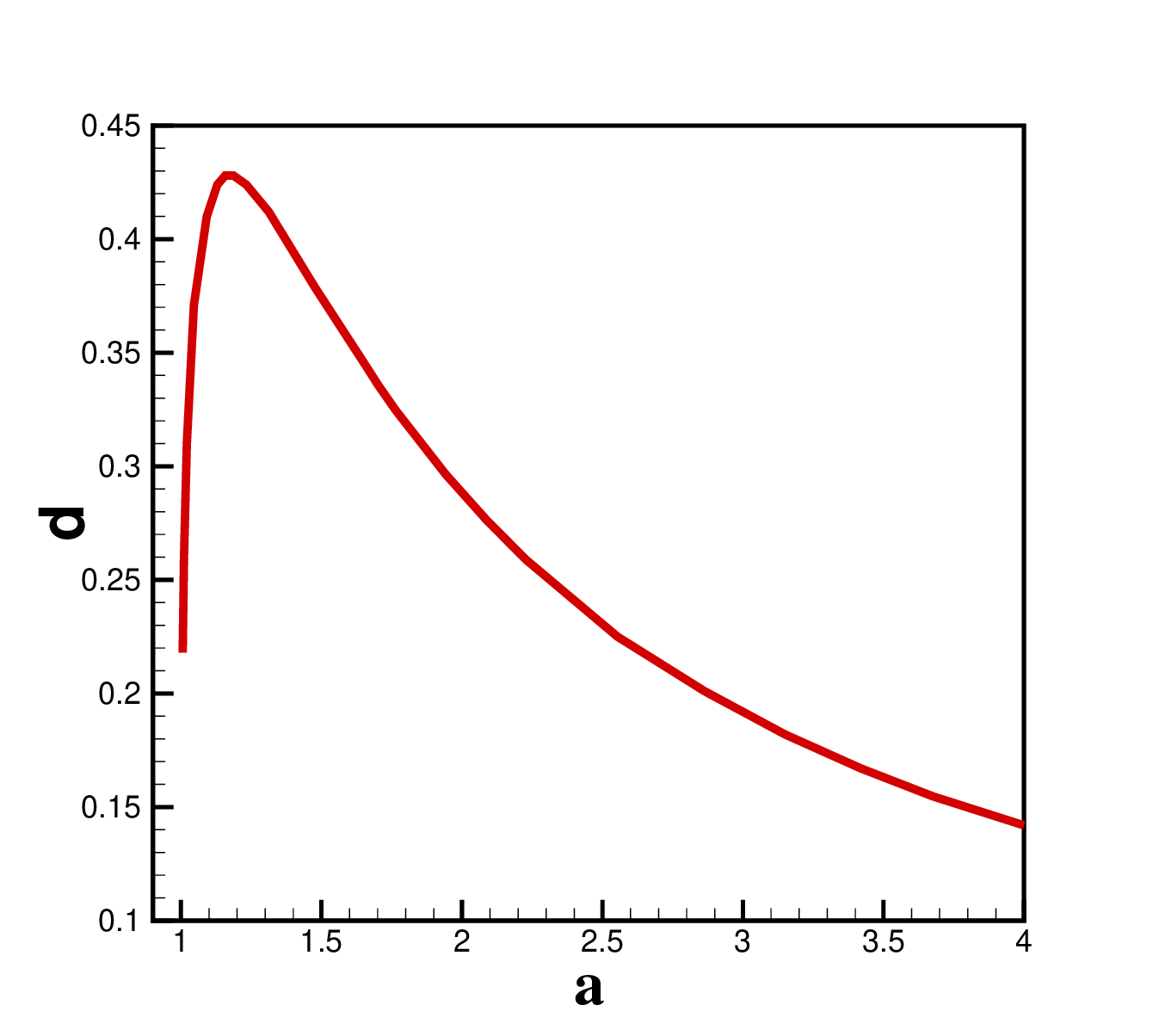} }
 \caption{Left: The radius of a rotating quark-antiquark at the boundary
 versus the angular velocity. Right: The radius of a rotating quark-antiquark at the boundary versus
  $a$.\label{d-w1}}
\end{figure}

By using numerical solutions,  a plot of $d$  as a function of
angular velocity $\omega$ is given in Fig. \ref{d-w1} where it is
seen that both $\omega$ and $d$ increase up to a maximum. As
$\omega$ increases and goes past its maximum the radius of the
rotating open string decreases. Hence there are two possible
U-shaped configurations representing two different values of
$\omega$. However, only the right region of the plots is physical
where to an increasing $\omega$ there corresponds a decreasing
radius. The configurations with larger values of $\omega$ have
shorter length
 and hence more energetically favorable. Therefore the branch to the left
of the maximum in Fig. \ref{d-w1}. is presumably unstable and will
decay to the right branch.
\begin{figure}
\centerline{ \includegraphics[width=3.5in]{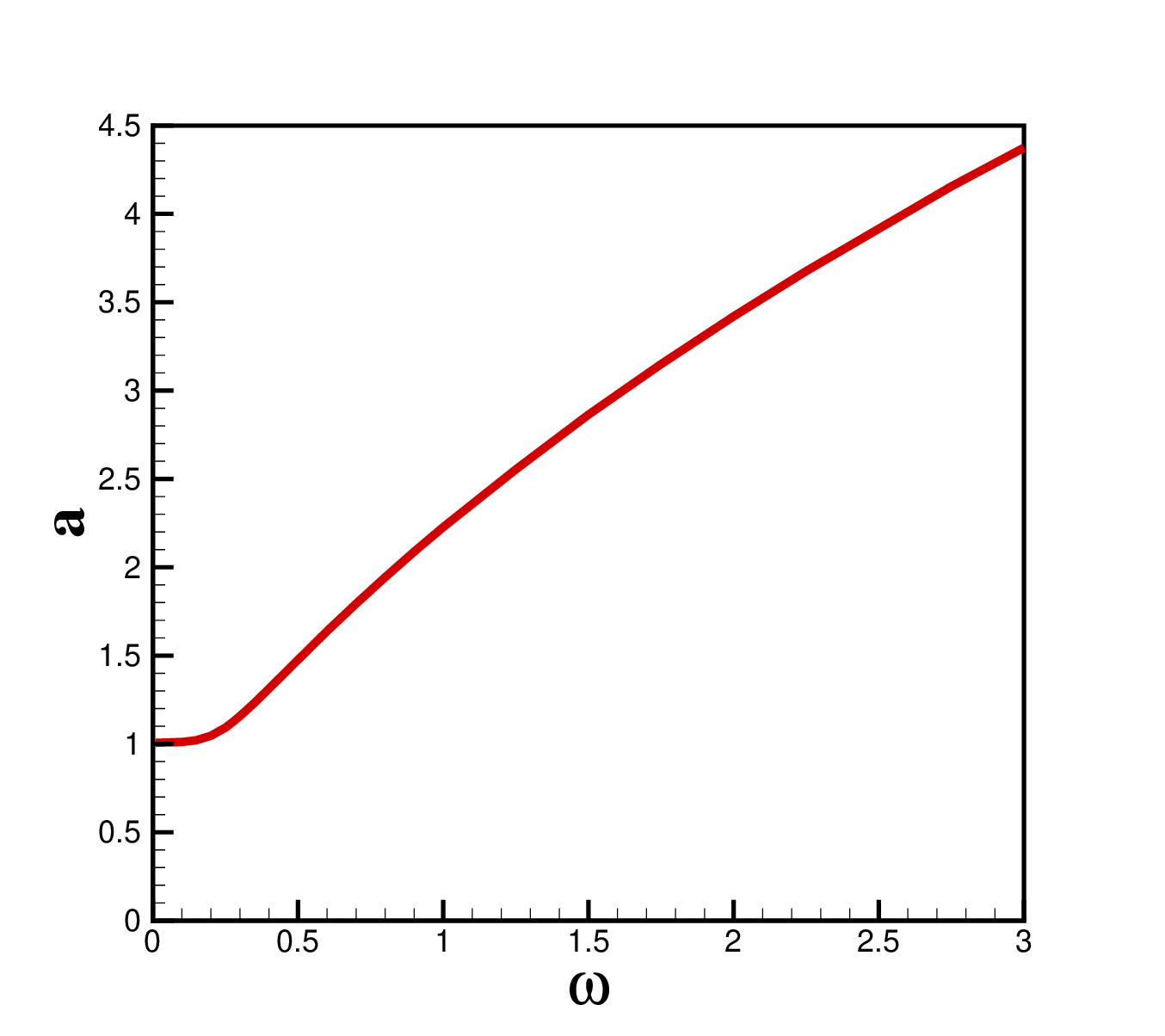}}
 \caption{We plot $a$ versus angular velocity $\omega$. \label{zminw}}
\end{figure}

It is interesting to investigate the behavior of a rotating
quark-antiquark pair in terms of the strings tip. To this end, we
introduce a new parameter $a=\frac{u_*}{u_h}$ where $u_*$ is the
nearest point of the (physical) string to the horizon. In the right
plot of the Fig. \ref{d-w1} we show $d$ versus $a$. In the physical
region, as $a$ increases the value of $d$ decreases. It is obvious
that there are two possible U-shaped string configurations at two
different values of $a$. The larger value of $a$ has shorter length
and hence it describes physical configurations. In order to check
the behavior of angular velocity $\omega$ in terms of $a$, we plot
them in Fig. \ref{zminw}. In the physical region, angular velocity
$\omega$ is approximately a linear
function of $a$. We fit this linear curve using %
\be\label{a-w} %
 a=m+n\omega+p\omega^2.
\ee %
A numerical analysis of \eqref{a-w} gives optimal values $m=0.70,
n=1.65$ and $p=-0.14$. The
$d$-$a$ curve is also fitted by%
\be\label{d-a2} %
 d(a)=\frac{s}{a}+\frac{r}{a^5}, %
\ee %
where $s=0.57$ and $r=-0.01$. The $d$-$\omega$ curve is easily found
by substituting \eqref{a-w} in \eqref{d-a2} %
\be\label{d-w12} %
 d(\omega)=\frac{0.57}{0.70+1.65\omega-0.14\omega^2}-\frac{0.01}{(0.70+1.65\omega-0.14\omega^2)^5},
\ee \label{d-weq}%
which reproduces the $d$-$\omega$ plot in Fig. \ref{d-w1}. It is
important to notice that $d$ and $a$ are always positive and thus we
have an upper limit on $\omega$. Without the third term in
\eqref{a-w}, $a$ and $d$ are always positive and $\omega$ is free to
go to infinity. We draw the reader's attention to the fact that in
the physical region $\omega$ cannot be zero.

In \cite{Rey} the heavy \emph{static} quark-antiquark potential as a
function of interquark distance $d$ was defined. It was argued that
at the particular point $d=d_*$ the potential curve crosses zero.
Gauge-string duality implies that the U-shaped string will break
into two straight open strings having no interaction energy and
vanishing potential for $d\geq d_*$. Furthermore, in the same paper
the relation between $d$ and $a$ was also considered. There, two
possible U-shaped string configurations at two different values of
$a$ arise. In the physical region, as $a$ increases the value of $d$
decreases. In
addition, for large $a$, $d$ was obtained as a function of $a$%
\be\label{a(d)} %
 d(a)=\frac{2c\lambda}{u_h}\left(\frac{1}{a}-\frac{1}{5a^5}-\frac{1}{10a^9}-...\right),
\ee %
where $c=0.56$.

\begin{figure}[ht]
\centerline{ \includegraphics[width=3.5in]{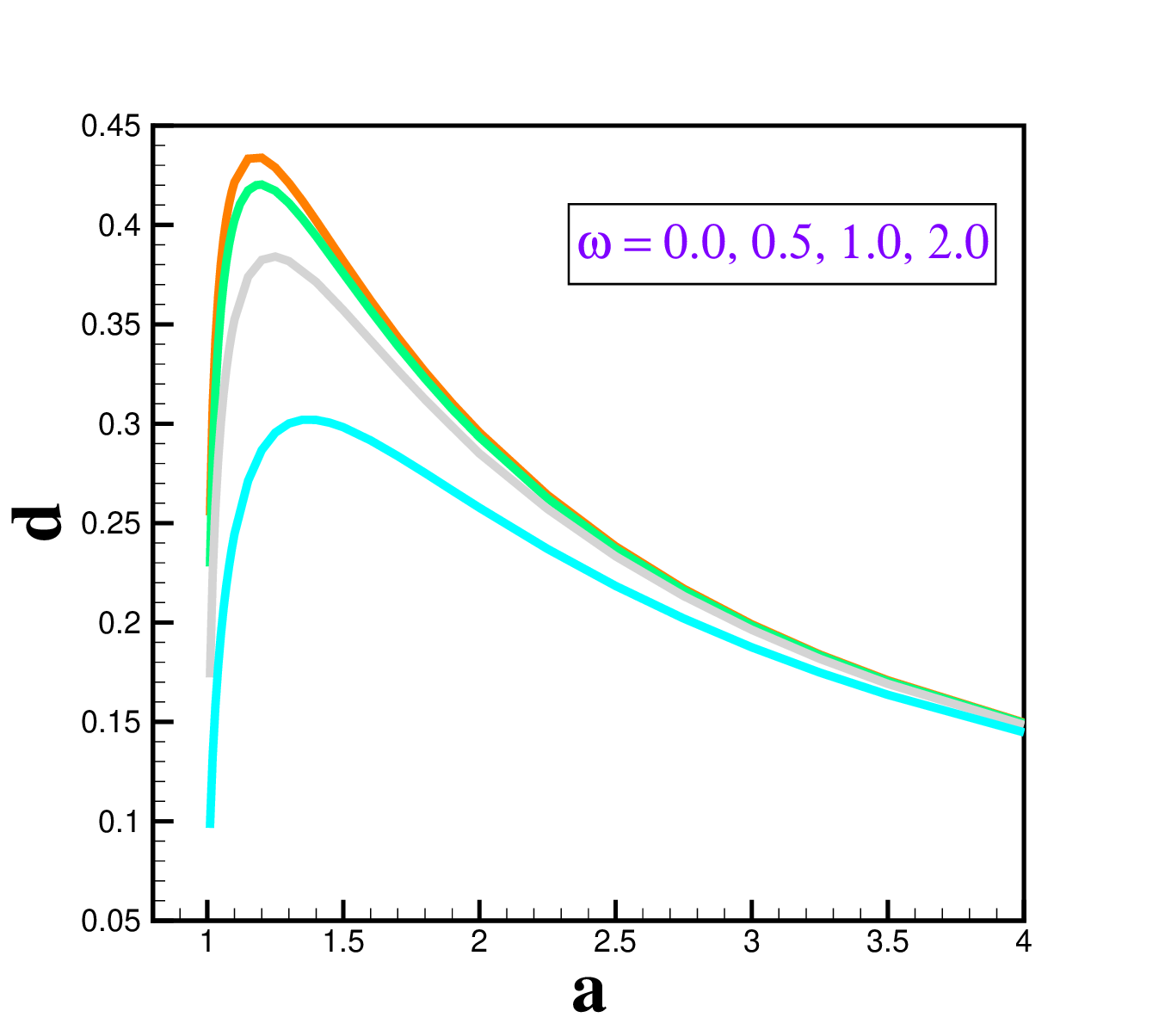}}
 \caption{We plot $d$ versus $a$ for different values of angular velocity of
the quark-antiquark pair. From maximum $d$ to minimum $d$, angular
velocity is $w=0.0,0.5,1.0,2.0$. In this figure we have fixed
angular velocity of the rotating string. Also we have not considered
Neumann boundary conditions (boundary is located at z=20). See the
similarity with rotating baryon in Figure 5 of
\cite{baryon}.\label{wfix}}
\end{figure}

In the our case we have plotted the radius of a rotating
quark-antiquark system in terms of $a$ and in comparison with static
case we find similar curves which we reproduce in Fig. \ref{wfix}.
One may conclude that the behavior of $d$ as a function of $a$ is
the same as in the static case \eqref{a(d)}, where the coefficients
depend on
angular velocity. In other words for large $a$ we may have %
\be\label{d-a} %
 d(a)=\frac{2c\lambda}{u_h}\left(\frac{\alpha(\omega)}{a}-\frac{\beta(\omega)}{5a^5}
 -\frac{\gamma(\omega)}{10a^9}-...\right),
\ee %
where %
\be %
 \alpha(\omega=0)=\beta(\omega=0)=\gamma(\omega=0)=1.
\ee %
It would be interesting to find values for $\alpha,\,\,\beta$ and
$\gamma$ \cite{working}.

Moreover it is clear that angular velocity is responsible for both
the variation of the radius of the quark-antiquark system as well as
$a$, see Fig. \ref{zminw}. Similarly to our discussion of the
potential in the static case, we find that in the \emph{rotating}
case for a particular value $\omega_*$ such that
$\omega_*>\omega_{max}$, corresponding to $d_*<d_{max}$,
dissociation appears. Once dissociation has appeared we are left
with a pair of straight strings whose potential vanishes for
$\omega<\omega_*$ ($d>d_*$).

A string-junction holographic model of a probe baryon in finite
temperature SYM dual to AdS-Schwarzschild black hole is studied in
\cite{Athanasiou:2008pz,baryon,Sin:2009dk}. Therein, a relation
between D5-brane position, which plays the same role of $a$ of the
U-shaped string in our work, and quark separation is plotted. It was
shown that quark separation becomes larger at first to turn smaller
as the position of the D5-brane increases. It was argued that in the
physical region by increasing the position of the D5-brane the quark
separation decreases. Interestingly, it was also discussed that to
an increase of the angular velocity there corresponds a decrease of
the maximum height of quark separation. Importantly, they did not
consider Neumann boundary conditions in this case. Since the results
match ours we may conclude that in our case the maximum height of
the radius of the quark-antiquark system depends on angular
velocity. On the other hand, the $\omega$ dependence of $\alpha$,
$\beta$ and $\gamma$ should control the height of the maximum
\cite{working}. In order to verify the above guess at least
numerically we do not consider boundary conditions and we plot in
Fig. \ref{wfix} quark-antiquark separation in terms of $a$ for
different angular velocities. It is easy to see that by increasing
angular velocity the maximum value of $d$ decreases.

\subsection{Dissociation of rotating quark-antiquark pair}

\begin{figure}[ht]
\centerline{\includegraphics[width=3.3in]{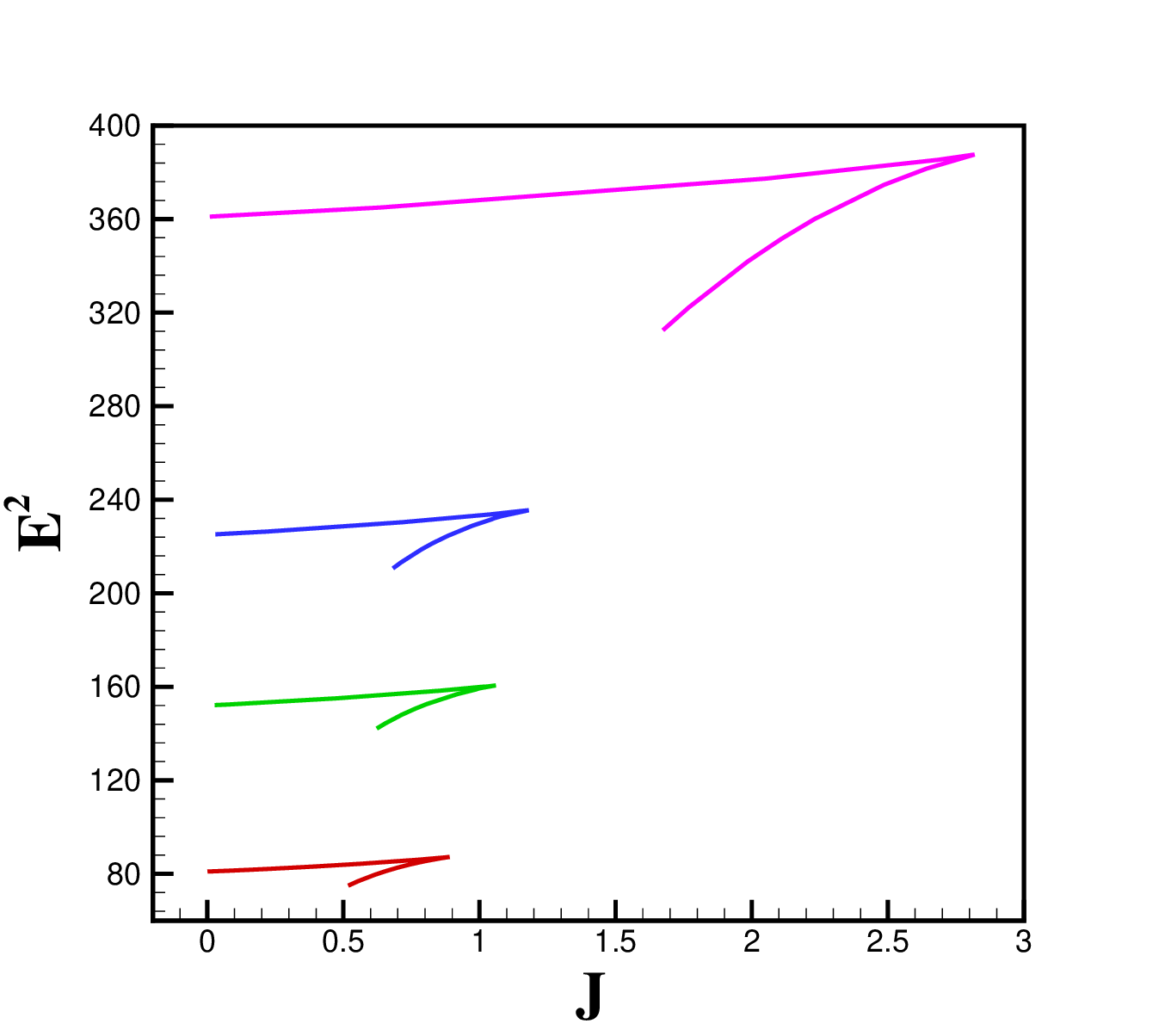}}%
\caption{The energy squared of a rotating string is plotted versus
angular momentum. From the upper curve to the lower curve,
temperature is increasing which corresponds to $z_h=1, 1.25, 1.50,
2.0$. \label{AdS5resultE2J}}
\end{figure}

 In this subsection, we analyze the dissociation of a rotating quark-antiquark
 pair. This subject has been studied in \cite{Peeters,Antipin}.
 As we found in the previous section angular velocity
 controls the radius, the length and the nearest point of the string to the horizon. We are going to
 understand the effect of rotation on the dissociation of a rotating quark-antiquark
 pair by studying the energy $E$ and angular momentum $J$. These are constants of motion and they can be
 easily found from the action (\ref{action}). The angular momentum of the quark-antiquark is
\be
 J=\omega\,\int d\rho \, \rho^2 z^4\left( \frac{
 \frac{z^{\prime2}}{z^4-1}+1}{\sqrt{-g} }\right).
\ee %
The other constant is the energy of the system that
follows from the Hamiltonian %
\be %
 {\cal{H}}=\dot{\theta} \frac{\partial
 {\cal{L}}}{\partial \dot{\theta}}-{\cal{L}},
\ee %
and it can be found to be  %
\be %
 E=\int d\,\rho \frac{z'^2+z^4-1}{\sqrt{-g}}.%
\ee %
It is not easy to find analytical solutions for $E(\omega)$ and
$J(\omega)$, so we resort to numerical analysis to obtain more
information about these quantities. The results are shown in Fig.
\ref{AdS5resultE2J} where we plotted the energy squared of the
rotating quark-antiquark pair in terms of its angular momentum.
Horizon and boundary are located at $z_h=1$ and $z=20$,
respectively. From the upper curve to the lower curve we see that
temperature increases. From equation (\ref{temperature}) it is clear
that a change in temperature corresponds to a change in the location
of the horizon. There is a maximum energy and angular velocity
beyond which the meson will dissociate.

At finite temperature, as $\omega$ decreases, the effective tension
of the string in the region near the horizon decreases. This leads
to the appearance of a maximum in energy and spin. It is natural to
interpret the temperature at which this happens as the critical
temperature at which a quark-antiquark of spin $J_{max}$ melts. Thus
we deduce that the dissociation temperature for quark-antiquark is
spin dependent. As temperature increases the maximum value of the
spin that a quark-antiquark can carry decreases, see Fig
\ref{AdS5resultE2J}. Moreover, we see that two states with identical
spin have different energies. The ones with smaller $\omega$ are
more energetic than the ones with larger $\omega$. This entails that
larger values of $\omega$ are more stable. We eliminate the variable
$\omega$ from the equations for energy and spin to obtain a plot of
energy versus spin shown in Fig. \ref{AdS5resultE2J}.

\section{Finite coupling corrections and rotating $q\bar{q}$ pair}
Let us now study finite coupling corrections to the rotating
quark-antiquark pair by considering ${\cal{R}}^2$ and ${\cal{R}}^4$
corrections in the dual gravity background. We should emphasize that
in the case of ${\cal{R}}^2$ correction, one cannot predict a result
for $\mathcal{N}=4$ SYM because the first higher derivative
correction in weakly curved type IIB backgrounds enters at order
${\cal{R}}^4$. These corrections on the drag force have been studied
in \cite{Fadafan}.

We follow the same procedure as in the previous section and start
with the Nambu-Goto action in these two different backgrounds. We
will see, as it is expected, results for ${\cal{R}}^4$ correction
are almost similar to AdS-Schwarzschild black hole without
correction studied in the previous section.

\subsection{${\cal{R}}^4$ corrections to AdS-Schwarzschild black brane}
We begin by studying a rotating string in the $\alpha'$ corrected
background. As before we will write the Nambu-Goto action and then
discuss the different aspects of rotating strings.

Since AdS/CFT correspondence refers to complete string theory, one
should consider the string corrections to the 10D supergravity
action. The first correction occurs at order $(\alpha')^3$
\cite{correction}. In the extremal $AdS_5\times S^5$ it is clear
that the metric does not change \cite{Banks}, conversely this is no
longer true in the non-extremal case. Corrections in inverse 't
Hooft coupling $1/\lambda$ which correspond to $\alpha^{\prime}$
corrections on the string theory side were found in \cite{alpha2}.
The $\alpha^{\prime}$-corrected metric is \cite{alpha1}

\be\label{metricform}%
 ds^2=G_{tt}\ dt^2+G_{xx}(d\rho^{2}+\rho^{2}d
\theta^{2}+dx_{3}^{2})+G_{uu}\ du^2, %
\ee %
where the metric functions are given by %
\begin{eqnarray}
G_{tt}&=&-R^{-2} u^2 (1-z^{-4})T(z),\nonumber\\
G_{xx}&=&R^{-2} u^2 X(z),\nonumber\\
G_{uu}&=&R^2 u^{-2}(1-z^{-4})^{-1} U(z),
\end{eqnarray}
and %
\begin{figure}[ht]
\centerline{\includegraphics[width=3.3in]{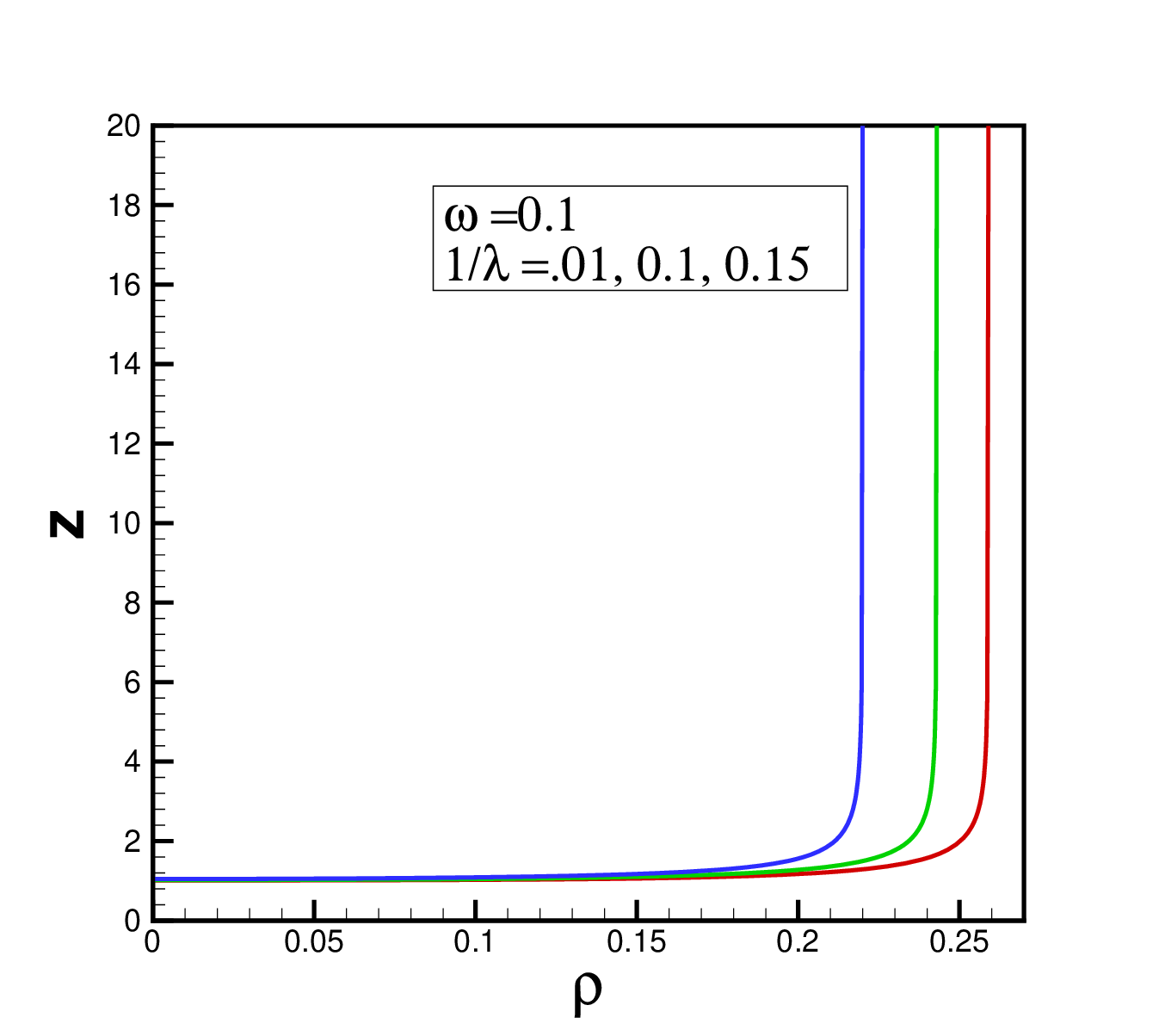}}%
\caption{Shape of a rotating string for different values of
$\lambda^{-1}$ at fixed angular velocity $\omega=0.1$. From right to
left $\lambda^{-1}=0.01,\,0.1\,,0.15$\label{R41}}
\end{figure}%
\begin{eqnarray}
T(z)&=&1-b\bigg(75z^{-4}+\frac{1225}{16}z^{-8}-\frac{695}{16}z^{-12}\bigg)+\dots ,\nonumber\\
X(z)&=&1-\frac{25b}{16}z^{-8}(1+z^{-4})+\dots,\nonumber\\
U(z)&=&1+b\bigg(75z^{-4}+\frac{1175}{16}z^{-8}-\frac{4585}{16}z^{-12}\bigg)+\dots\
.
\end{eqnarray}
There is an event horizon at $u=u_h$ and the geometry is
asymptotically AdS at large $u$ with a radius of curvature $R$.
The expansion parameter $b$ can be expressed in terms of the
inverse 't Hooft coupling as
\be\label{blambda}%
 b=\frac{\zeta(3)}{8}\lambda^{-3/2}\sim .15\lambda^{-3/2}.
\ee %
The dynamics of a rotating string in this background is described by
the Nambu-Goto action
\begin{eqnarray}\label{max6} %
 S&=&-\frac{u_h}{2\pi}\int
 dt\,d\rho\sqrt{\bigg((z^4-1)T(z)-\rho^2 \omega^2 z^4X(z)\bigg)
\left(\frac{z^{\prime2}}{z^4-1}U(z)+X(z)\right )},
\end{eqnarray} %
where, in deriving \eqref{max6}, we used \eqref{ansatz} and \eqref{changeofvariable}.
The positivity condition becomes %
\be %
 (z^4-1)T(z)\geq\rho^2 \omega^2 z^4X(z). %
\ee %
One can easily find equation of motion for $z$ and then solve it
numerically leading to Fig. \ref{R41}. In this figure, for fixed
$\omega$, a plot of $z(\rho)$ versus $\rho$ is depicted. By
analyzing the shapes of the string for various values of
$\lambda^{-1}$ we see that as $\lambda^{-1}$ decreases the
separation of the string endpoints increases.

\begin{figure}[ht]
\centerline{\includegraphics[width=3in]{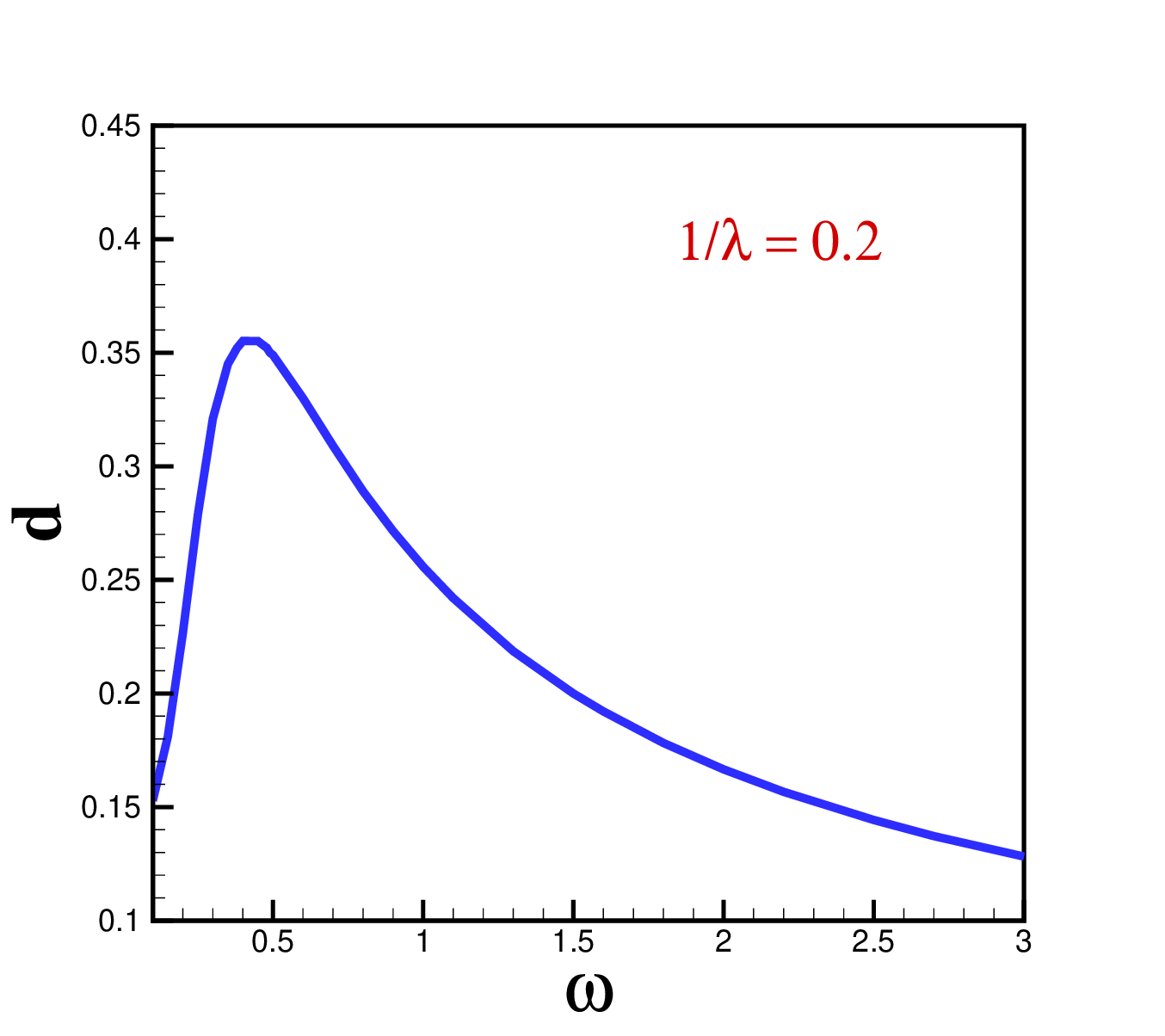}}%
\caption{The radius of a rotating meson versus the angular velocity
for fixed coupling, $\lambda ^{-1}=0.2$ .\label{1-Lambda.2}}
\end{figure}

For fixed $\lambda^{-1}$, as shown in Fig. \ref{1-Lambda.2}, we
again have physical and unphysical strings. For physical strings, as
angular velocity decreases the rotating string endpoints become more
separated and the U-shaped string penetrates deeper into the
horizon. One can compare Fig. \ref{1-Lambda.2} with previous results
in the case of no corrections. It is clear that the distance between
rotating quark and antiquark is smaller when
${\cal{R}}^4$ correction is considered.\\
It would be interesting to connect this result to experiments at
RHIC and LHC. Thermal properties of QGP can be investigated by
quarkonium states. The suppression of heavy quark bound states in
RHIC is a sensitive probe for QGP. Quarkonium is a bound state of a
heavy quark and antiquark pair and it is possible to introduce
screening length for it. When the screening radius is smaller than
the radius of the bound state quarkonium will dissociate. Then by
studying this physical quantity one can find the effect of finite
coupling corrections on the dissociation of charmonium and
bottominium at RHIC. This can be done in detail by studying the
potential of the rotating quark-antiquark and we leave this problem
for further study \cite{working}.

Since the ${\cal{R}}^4$ correction has a small effect on other
results of the previous section  we do not repeat them.%
\subsection{Gauss-Bonnet gravity background}
We continue our analysis and consider $\mathcal{R}^2$ correction. On
the gravity side it corresponds to finite coupling correction on the
gauge theory side. Motivation to consider this correction comes from
the fact that string theory contains higher derivative corrections
from stringy ($1/\lambda$) or quantum effects ($1/N$). In other
words, a possible violation of the lower bound of the ratio of shear
viscosity $\eta$ to entropy density $s$ in GB background was shown
in \cite{Kats:2007mq,Brigante:2008gz}, also see \cite{Buchel:2008vz}
and references therein. Following this observation, we study
properties of rotating quark-antiquark in GB gravity.

In five dimensions, the most general theory of gravity with
quadratic powers of curvature and exact solution is
Einstein-Gauss-Bonnet (EGB) theory which is defined by the following action \cite{Zwiebach} %
\be %
 S=\frac{1}{16\pi G_N}\int d^5x\sqrt{-g}\left({\cal{R}}+\frac{12}{R^2}+\frac{\lambda_{GB} R^2}{2}({\cal{R}}^2
 -4{\cal{R}}_{\mu\nu}{\cal{R}}^{\mu\nu}+{\cal{R}}^{\mu\nu\alpha\beta}{\cal{R}}_{\mu\nu\alpha\beta})\right).
\ee %
Here $R$ is proportional to the radius of the asymptotic AdS space
and $\mu$ runs over 1 to 5. We are going to consider how curvature
squared terms affect rotating
strings. There is an exact black brane solution whose metric is of the form \eqref{metricform} with \cite{GB}%
\be %
 G_{tt}=-kR^{-2}u^2h(u),\ \ \ G_{xx}=R^{-2}u^2,\ \ \
 G_{uu}=R^2u^{-2}h^{-1}(u),
\ee %
where
\begin{equation}
h(u)= \frac{1}{2\lambda_{GB}}\left[ 1-\sqrt{1-4 \lambda_{GB}\left(
1-\frac{u_h^4}{u^4} \right)}\ \right],\ \ \
k=\frac12(1+\sqrt{1-4\lambda_{GB}}).
\end{equation}
The scaling factor $k$ for $G_{tt}$ ensures us that the speed of
light at the boundary theory is one. Beyond
$\lambda_{GB}\leq\frac{1}{4}$ there is no vacuum AdS solution and
one cannot have a conformal field theory at the boundary. New bounds
for $\lambda_{GB}$ come from causality condition. Authors of
\cite{Brigante:2008gz} found that in five dimensions
$\lambda_{GB}\leq 0.09$ and when dimensions of spacetime go up,
causality restricts the value of $\lambda_{GB}$ in the region
$\lambda_{GB}\leq 0.25$ . The Hawking temperature of the black hole
is given by%
\be %
 T=\frac{\sqrt{k}\ u_h}{\pi R^2}.
\ee %
Using \eqref{ansatz} and \eqref{changeofvariable}, the Nambu-Goto
action becomes %
\begin{eqnarray} %
 S&=&-\frac{u_h}{2\pi}\int
 dtd\rho\sqrt{(kh(z)-\rho^2\omega^2)
 \left(z^4+z^{\prime2}h^{-1}(z)\right)}\nonumber\\
 &\equiv&-\frac{u_h}{2\pi} \int
 dtd\rho {\cal{L}},
\end{eqnarray} %
where %
\be %
 h(z)= \frac{1}{2\lambda_{GB}}\left[ 1-\sqrt{1-4
 \lambda_{GB}\left( 1-z^{-4}\right)}\ \right]. %
\ee %
Positivity of the square root in the action leads to %
\be\label{max2} %
 u(\rho)\geq\frac{u_h}{\bigg(1-\frac{1}{4\lambda_{GB}}\Big(1-(1-\frac{2\lambda_{GB}}{k}\rho^2\omega^2)^2\Big)\bigg)^{1/4}}.
\ee %
\begin{figure}[ht]
 \includegraphics[width=3in]{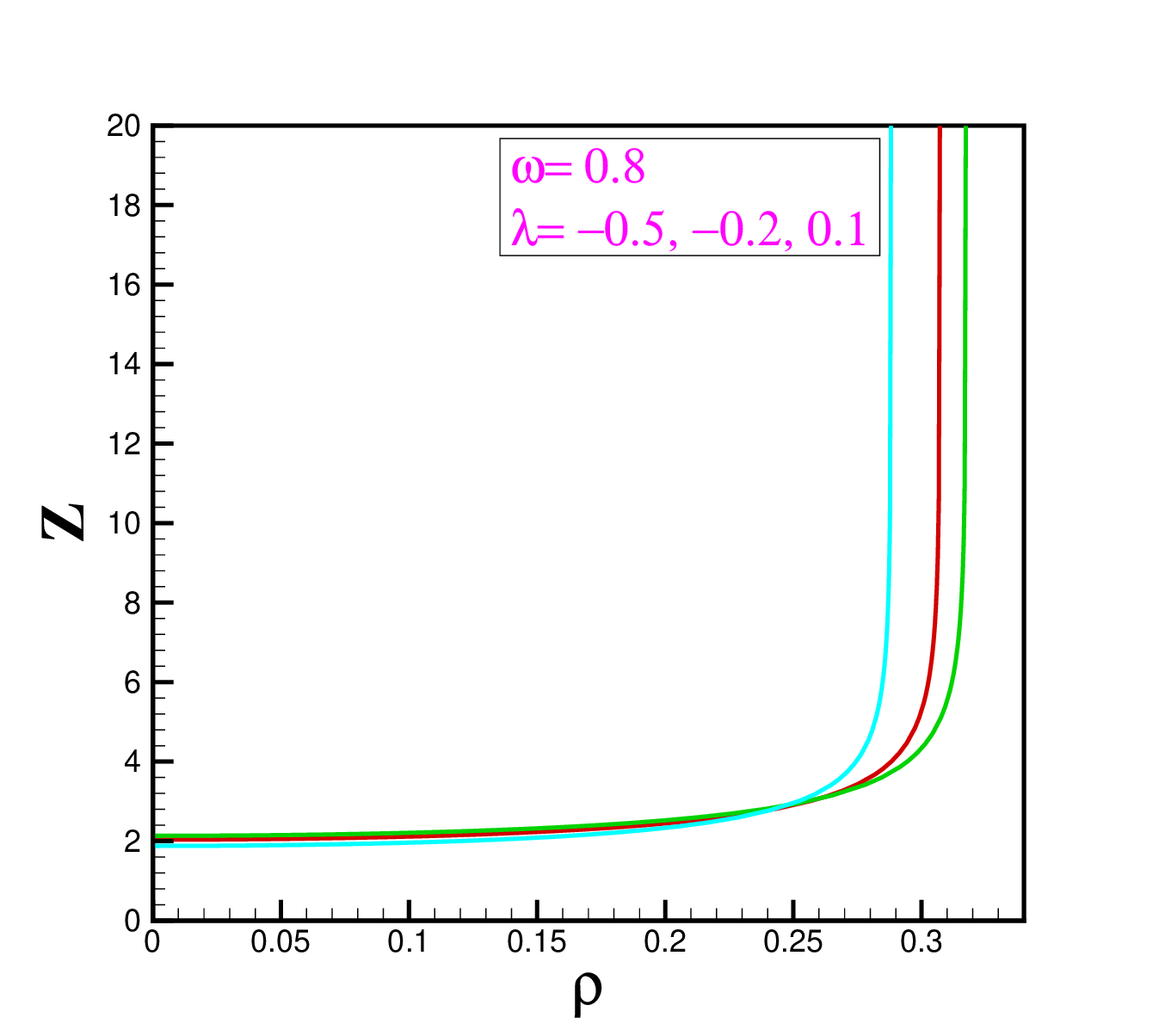}\includegraphics[width=3in]{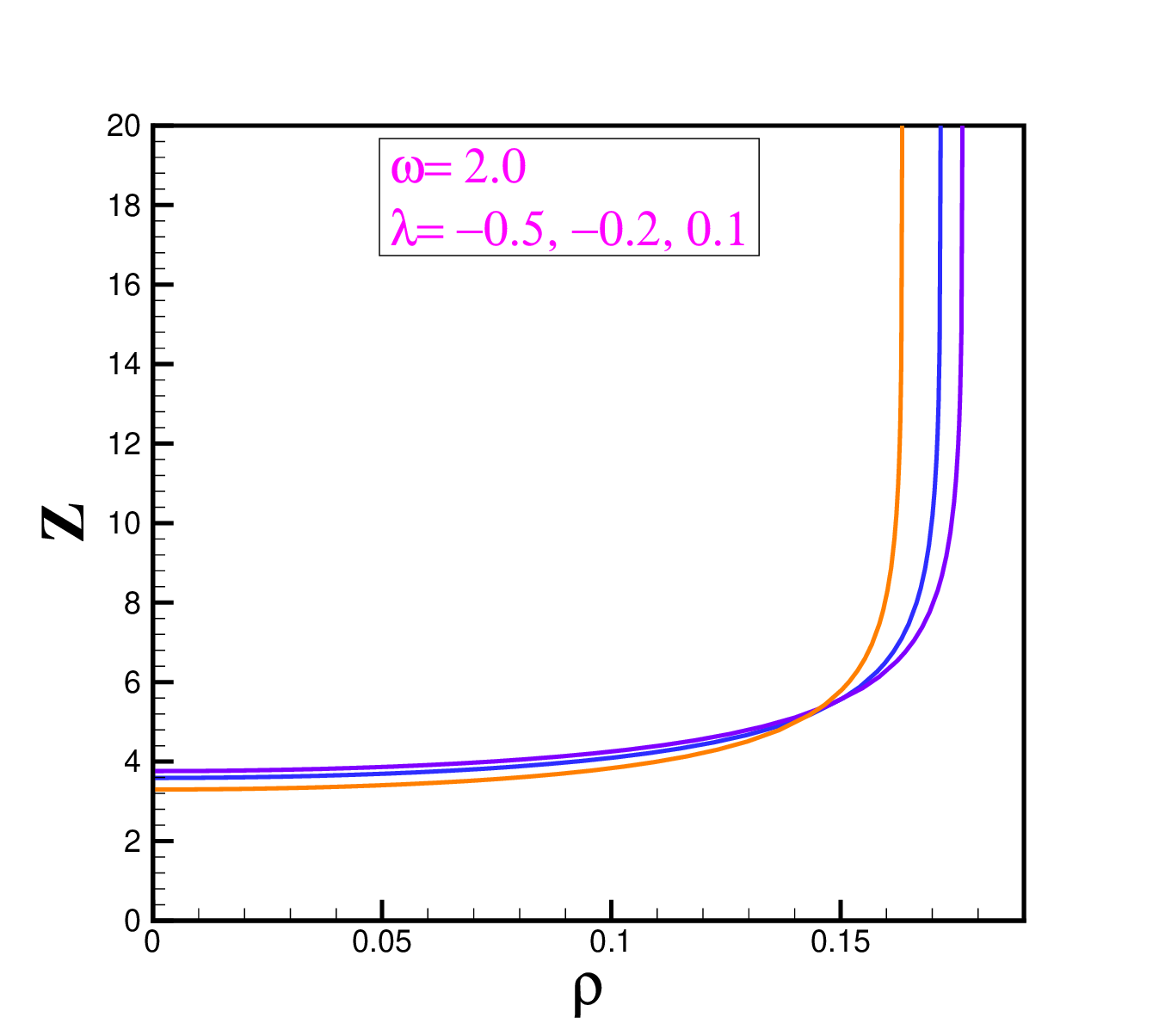}
 \caption{Shapes of a rotating heavy meson for different values of coupling
 constants.
 Left: For $\omega=0.8$ coupling constants $\lambda_{GB}=-0.5, -0.2, 0.1$ from the right to the left curve.
 Right: For $\omega=2.0$ coupling constants $\lambda_{GB}=-0.5, -0.2, 0.1$ from the right to the left curve.\label{GB1}}
\end{figure}
Equation \eqref{max2} implies that rotating strings have to lie
above this curve. The equation of motion for $z$ is %
\be\label{max7} %
 \partial_\rho\left(\frac{z'h^{-1}(z)\big(kh(z)-\rho^2\omega^2\big)}
 {\sqrt{-g}}\right)-\frac{\partial{\cal{L}}}{\partial z}=0.
\ee %
One can solve this equation numerically to obtain more information
about a rotating quark-antiquark pair in the EGB background.
Boundary conditions which solve the differential equation
\eqref{max7} are $z(d)=20$ and $z'(d)\rightarrow \infty $. Also the
condition $z'(0)=0$ at the tip of the string is employed. The shapes
of rotating open string, for fixed temperature and different angular
velocities and $\lambda_{GB}$ are shown in Fig. \ref{GB1}. By
analyzing the shapes of the string for different values of
$\lambda_{GB}$ with fixed $\omega$, as $\lambda_{GB}$ increases the
string endpoints become more separated {\em{i.e.}} the radius $d$ of
the rotating open string at the boundary increases but the tip of
the U-shaped string does not change considerably. Also it is evident
that shorter strings, {\em{i.e.}} smaller $d$, have larger angular
velocities which similarly to our previous cases. Comparing the two
plots in Fig. \ref{GB1}, we realize that for larger angular
velocities there corresponds smaller radii of the quark-antiquark
system.

\subsubsection{Energy and spin}
The spin and energy of a quark-antiquark in the above background
are easily defined by %
\be\begin{split} %
 J&=\omega\, \int d\rho\,\rho^2 \left(\frac{z^4+z^{\prime2}h^{-1}(z)}{\sqrt{-g}}\right),\cr
 E&=k\,\int d\rho
 \frac{h(z)\big(z^4+z^{\prime2}h^{-1}(z)\big)}{\sqrt{-g}}.
\end{split}\ee %
\begin{figure}[ht]
\centerline{\includegraphics[width=4in]{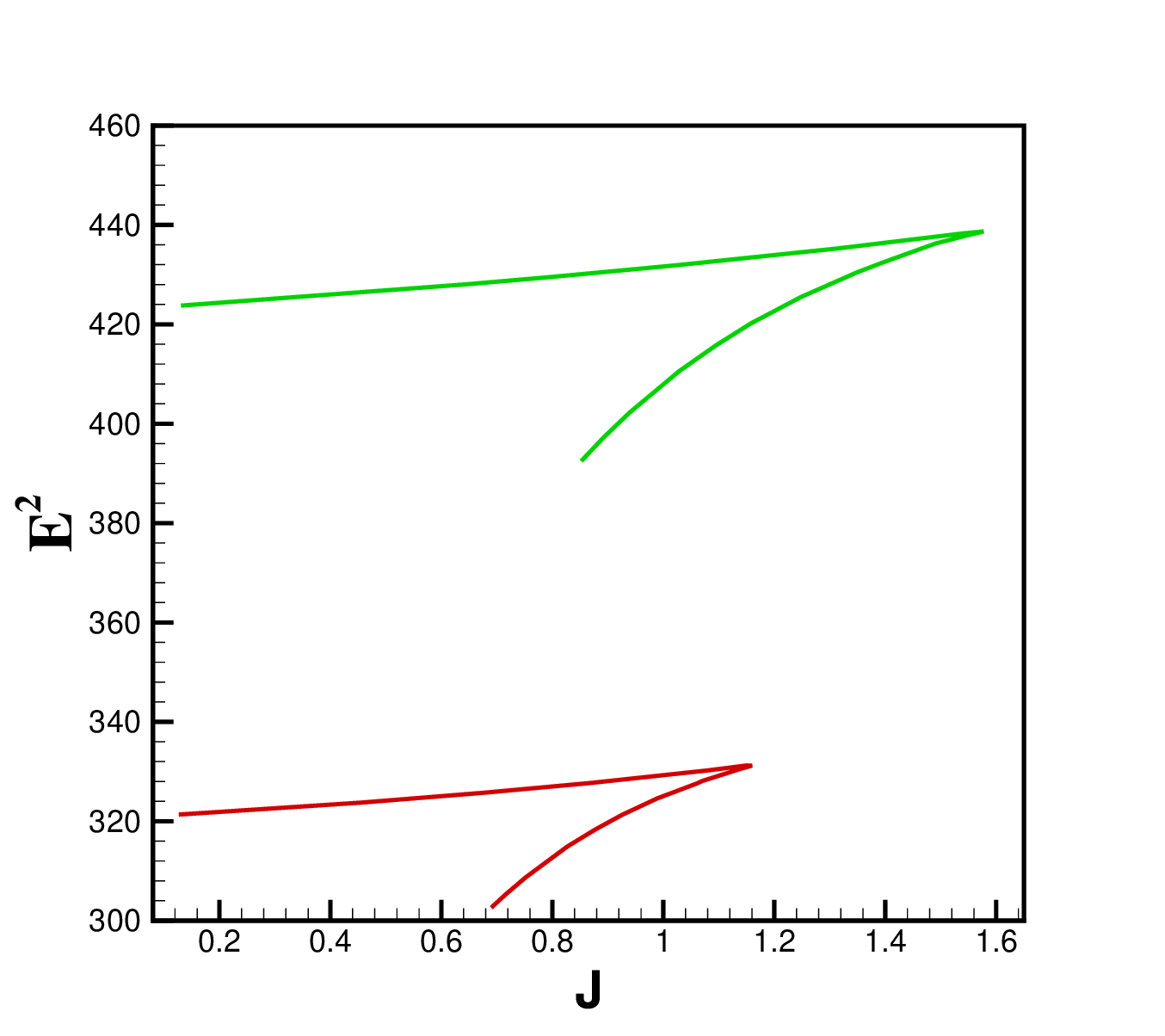}}%
\caption{The energy squared of rotating string in terms of the spin
in hot plasma dual to Gauss-Bonnet gravity. $u_h=1$ and boundary is
located at $u=20$. For upper curve $\lambda_{GB}=-0.2$ and for lower
curve $\lambda_{GB}=+0.1$. It is clear that there is a maximum
energy and spin where beyond them the meson will be
dissociated.\label{GBE2J}}
\end{figure}%
Energy squared in terms of spin is plotted in Fig. \ref{GBE2J}. It
is evident that like in previous cases there is a maximum which may
be related to a dissociation point. The maximum of the energy
squared is noticeably enhanced respect to prior cases. Also, the
effect of the GB coupling is seen in the same figure whereby an
increase of $\lambda_{GB}$ yields a decrease of the maximum energy
and spin. This is an interesting phenomenon that shows us the role
of finite coupling corrections. If we compare it with the case of no
corrections, a rotating meson will dissociate with smaller energy
and spin.

\section{Conclusion}
In this paper we studied a rotating quark-antiquark in
AdS-Schwarzschild black hole and solved its equation of motion
numerically. By using suitable boundary conditions, at fixed
temperature, we find two U-shaped strings of which the shorter one
is energetically stable. By analyzing string shapes, corresponding
to physical regions, we deduce that as angular velocity decreases
({\em{i.e.}} $a$ decreases) the string endpoints become more
separated, that is the radius of the rotating string at the boundary
increases. Interestingly, our results match with a similar analysis
carried out in Sakai-Sugimoto model \cite{Peeters}. In addition, in
\eqref{a-w} we were able to cast an equation which expresses the
distance between the tip of the rotating open string and the horizon
in terms of angular velocity. The upper limit on angular velocity
comes from the positivity of $d$.

Our study has lead us to equation \eqref{d-a} which describes the
radius of a rotating quark-antiquark in terms of $a$ and $\omega$
where remarkably the dependence on $a$ is exactly as for the static
case. We plan to derive the $\alpha, \beta$ and $\gamma$ functions
appearing in \eqref{d-a} in a future work \cite{working}. This can
be done by either best fitting methods or analytically as in
\cite{Rey}. We also found a similarity between our setting and the
baryonic case. We showed that the tip of the string in our model and
the position of a D5-brane in the baryonic case play the same role:
by increasing angular velocity both of them behave alike.

By plotting energy in terms of spin we showed that at a fixed
temperature spin has a maximum value interpreted as a
temperature-dependent dissociation spin.

We also extended our setup as to include ${\cal{R}}^2$ and
${\cal{R}}^4$ corrections. We summarize our results as follows:%
\begin{itemize}
    \item The effects of ${\cal{R}}^4$ higher order  correction can be
controlled by $\lambda^{-1}$. For a fixed $\omega$, by increasing
$\lambda^{-1}$ the radius of the rotating open string increases
although the tip of the string is almost constant. In other words,
in presence of higher order correction the rotating open string has
longer length. Since the longer string is more energetic we conclude
that the ${\cal{R}}^4$ correction increases the value of energy. In
short, the radius, length and energy of a rotating open string will
increase when ${\cal{R}}^4$ correction becomes bigger. We showed
that ${\cal{R}}^2$ correction yields the same results as
${\cal{R}}^4$ correction does.
    \item As the potential between a quark and antiquark becomes zero,
    a free quark and antiquark are produced. In the case of a rotating open string
it happens when $d=d_*$($\omega=\omega_*$). Because of ${\cal{R}}^4$
correction the value of $\omega_*$ decreases which indicates that
the quark-antiquark system can be separated more easily than before.
    \item We considered ${\cal{R}}^2$ correction in GB-theory and found
that there is a sensible enhancement in energy.
\end{itemize}
We conclude with a final remark. Finite coupling corrections may be
useful to study the dissociation of quarkonium at RHIC. From this
point of view our results can be interpreted as melting of
quarkonium states.

\section*{Acknowledgment}
We would like to thank  J. Sonnenschein and K. Hashimoto for their
helpful comments and specially thank M. Edalati for reading the
manuscript and valuable discussion. It is also a pleasure to thank
M. Vincon for reading the manuscript very carefully.


\begin{thebibliography}{50}
\bibitem{MAGO}
  O.~Aharony, S.~S.~Gubser, J.~M.~Maldacena, H.~Ooguri and Y.~Oz,
  ``Large N field theories, string theory and gravity,''
  Phys.\ Rept.\  {\bf 323}, 183 (2000)
  [arXiv:hep-th/9905111];

\bibitem{Maldacena:1997re}
  J.~M.~Maldacena,
  ``The large N limit of superconformal field theories and supergravity,''
  Adv.\ Theor.\ Math.\ Phys.\  {\bf 2} (1998) 231
  [Int.\ J.\ Theor.\ Phys.\  {\bf 38} (1999) 1113]
  [arXiv:hep-th/9711200]; S.~J.~Rey and J.~T.~Yee,
  ``Macroscopic strings as heavy quarks in large N gauge theory and  anti-de
  Sitter supergravity,''
  Eur.\ Phys.\ J.\  C {\bf 22} (2001) 379
  [arXiv:hep-th/9803001].


\bibitem{Gubser:1998bc}
  S.~S.~Gubser, I.~R.~Klebanov and A.~M.~Polyakov,
  ``Gauge theory correlators from non-critical string theory,''
  Phys.\ Lett.\  B {\bf 428} (1998) 105
  [arXiv:hep-th/9802109];

\bibitem{Witten:1998qj}
  E.~Witten,
  ``Anti-de Sitter space and holography,''
  Adv.\ Theor.\ Math.\ Phys.\  {\bf 2} (1998) 253
  [arXiv:hep-th/9802150];

\bibitem{Witten:1998zw}
  E.~Witten,
  ``Anti-de Sitter space, thermal phase transition, and confinement in  gauge
  theories,''
  Adv.\ Theor.\ Math.\ Phys.\  {\bf 2} (1998) 505
  [arXiv:hep-th/9803131];

\bibitem{Peeters:2007ab}
  K.~Peeters and M.~Zamaklar,
  ``The string/gauge theory correspondence in QCD,''
  Eur.\ Phys.\ J.\ ST {\bf 152} (2007) 113
  [arXiv:0708.1502 [hep-ph]].
\bibitem{Mateos:2007ay}
  D.~Mateos,
  ``String Theory and Quantum Chromodynamics,''
  Class.\ Quant.\ Grav.\  {\bf 24} (2007) S713
  [arXiv:0709.1523 [hep-th]].



\bibitem{Edelstein:2009iv}
  J.~D.~Edelstein, J.~P.~Shock and D.~Zoakos,
  ``The AdS/CFT Correspondence and Non-perturbative QCD,''
  AIP Conf.\ Proc.\  {\bf 1116} (2009) 265
  [arXiv:0901.2534 [hep-ph]].
\bibitem{Myers:2008fv}
  R.~C.~Myers and S.~E.~Vazquez,
  ``Quark Soup al dente: Applied Superstring Theory,''
  Class.\ Quant.\ Grav.\  {\bf 25} (2008) 114008
  [arXiv:0804.2423 [hep-th]].

\bibitem{Gubser:2009fc}
  S.~S.~Gubser,
  ``Using string theory to study the quark-gluon plasma: progress and perils,''
  arXiv:0907.4808 [hep-th].


\bibitem{Shuryak:2004cy}
  E.~V.~Shuryak,
  ``What RHIC experiments and theory tell us about properties of  quark-gluon
  Nucl.\ Phys.\  A {\bf 750} (2005) 64
  [arXiv:hep-ph/0405066];
  K.~Adcox {\it et al.}  [PHENIX Collaboration],
  ``Formation of dense partonic matter in relativistic nucleus
  collisions at RHIC: Experimental evaluation by the PHENIX  collaboration,''
  Nucl.\ Phys.\  A {\bf 757} (2005) 184
  [arXiv:nucl-ex/0410003];
  I.~Arsene {\it et al.}  [BRAHMS Collaboration],
  ``Quark gluon plasma and color glass condensate at RHIC? The perspective
  from the BRAHMS experiment,''
  Nucl.\ Phys.\  A {\bf 757} (2005) 1
  [arXiv:nucl-ex/0410020];
  J.~Adams {\it et al.}  [STAR Collaboration],
  ``Experimental and theoretical challenges in the search for the quark  gluon
  plasma: The STAR collaboration's critical assessment of the  evidence from
  RHIC collisions,''
  Nucl.\ Phys.\  A {\bf 757} (2005) 102
  [arXiv:nucl-ex/0501009];

\bibitem{Liu:2006nn}
  H.~Liu, K.~Rajagopal and U.~A.~Wiedemann,
  ``An AdS/CFT calculation of screening in a hot wind,''
  Phys.\ Rev.\ Lett.\  {\bf 98} (2007) 182301
  [arXiv:hep-ph/0607062].
\bibitem{Liu:2006ug}
  H.~Liu, K.~Rajagopal and U.~A.~Wiedemann,
  ``Calculating the jet quenching parameter from AdS/CFT,''
  Phys.\ Rev.\ Lett.\  {\bf 97} (2006) 182301
  [arXiv:hep-ph/0605178].
\bibitem{Herzog:2006gh}
  C.~P.~Herzog, A.~Karch, P.~Kovtun, C.~Kozcaz and L.~G.~Yaffe,
  ``Energy loss of a heavy quark moving through N = 4 supersymmetric
  Yang-Mills plasma,''
  JHEP {\bf 0607} (2006) 013
  [arXiv:hep-th/0605158].
\bibitem{CasalderreySolana:2006rq}
  J.~Casalderrey-Solana and D.~Teaney,
  ``Heavy quark diffusion in strongly coupled N = 4 Yang Mills,''
  Phys.\ Rev.\  D {\bf 74} (2006) 085012
  [arXiv:hep-ph/0605199].
\bibitem{Gubser:2006bz}
  S.~S.~Gubser,
  ``Drag force in AdS/CFT,''
  Phys.\ Rev.\  D {\bf 74} (2006) 126005
  [arXiv:hep-th/0605182].
\bibitem{Gubser:2006qh}
  S.~S.~Gubser,
  ``Comparing the drag force on heavy quarks in N = 4 super-Yang-Mills theory
  and QCD,''
  Phys.\ Rev.\  D {\bf 76} (2007) 126003
  [arXiv:hep-th/0611272].

\bibitem{Erdmenger:2007cm}
  J.~Erdmenger, N.~Evans, I.~Kirsch and E.~Threlfall,
  ``Mesons in Gauge/Gravity Duals - A Review,''
  Eur.\ Phys.\ J.\  A {\bf 35} (2008) 81
  [arXiv:0711.4467 [hep-th]].

\bibitem{de Forcrand:2000jx}
  P.~de Forcrand {\it et al.}  [QCD-TARO Collaboration],
  ``Meson correlators in finite temperature lattice QCD,''
  Phys.\ Rev.\  D {\bf 63} (2001) 054501
  [arXiv:hep-lat/0008005].

\bibitem{Gubser:2009md}
  S.~S.~Gubser and A.~Karch,
  ``From gauge-string duality to strong interactions: a Pedestrian's Guide,''
  arXiv:0901.0935 [hep-th].


\bibitem{Maldacena:1998im}
  J.~M.~Maldacena,
  ``Wilson loops in large N field theories,''
  Phys.\ Rev.\ Lett.\  {\bf 80} (1998) 4859
  [arXiv:hep-th/9803002].
\bibitem{Rey}
  S.~J.~Rey, S.~Theisen and J.~T.~Yee,
  ``Wilson-Polyakov loop at finite temperature in large N gauge theory and
  anti-de Sitter supergravity,''
  Nucl.\ Phys.\  B {\bf 527}, 171 (1998)
  [arXiv:hep-th/9803135];

\bibitem{Chernicoff:2006hi}
  M.~Chernicoff, J.~A.~Garcia and A.~Guijosa,
  ``The energy of a moving quark-antiquark pair in an N = 4 SYM plasma,''
  JHEP {\bf 0609} (2006) 068
  [arXiv:hep-th/0607089].
\bibitem{Argyres:2006vs}
  P.~C.~Argyres, M.~Edalati and J.~F.~Vazquez-Poritz,
  ``No-drag string configurations for steadily moving quark-antiquark pairs in
  a thermal bath,''
  JHEP {\bf 0701} (2007) 105
  [arXiv:hep-th/0608118].
\bibitem{Sadeghi:2008ci}
  J.~Sadeghi, B.~Pourhassan and S.~Heshmatian,
  ``Drag Force on Rotating Quark-Antiquark Pair in a N=4 SYM plasma,''
  arXiv:0812.4816 [hep-th].

\bibitem{Kruczenski:2003be}
  M.~Kruczenski, D.~Mateos, R.~C.~Myers and D.~J.~Winters
  ``Meson spectroscopy in AdS/CFT with flavour,''
  JHEP {\bf 0307} (2003) 049
  [arXiv:hep-th/0304032];
\bibitem{Kruczenski:2004me}
  M.~Kruczenski, L.~A.~P.~Zayas, J.~Sonnenschein and D.~Vaman,
  ``Regge trajectories for mesons in the holographic dual of large-N(c)  QCD,''
  JHEP {\bf 0506} (2005) 046
  [arXiv:hep-th/0410035];
\bibitem{Peeters}
  K.~Peeters, J.~Sonnenschein and M.~Zamaklar,
  ``Holographic melting and related properties of mesons in a quark gluon
  plasma,''
  Phys.\ Rev.\  D {\bf 74} (2006) 106008
  [arXiv:hep-th/0606195];
\bibitem{Antipin}
  O.~Antipin, P.~Burikham and J.~Li,
  ``Effective Quark Antiquark Potential in the Quark Gluon Plasma from
  Gravity Dual Models,''
  JHEP {\bf 0706} (2007) 046
  [arXiv:hep-ph/0703105];
  P.~Burikham and J.~Li,
  ``Aspects of the screening length and drag force in two alternative gravity
 duals of the quark-gluon plasma,''
  JHEP {\bf 0703}, 067 (2007)
  [arXiv:hep-ph/0701259];

\bibitem{Fadafan:2008bq}
  K.~B.~Fadafan, H.~Liu, K.~Rajagopal and U.~A.~Wiedemann,
  ``Stirring Strongly Coupled Plasma,''
  Eur.\ Phys.\ J.\  C {\bf 61} (2009) 553
  [arXiv:0809.2869 [hep-ph]].



\bibitem{Friess:2006rk}
  J.~J.~Friess, S.~S.~Gubser, G.~Michalogiorgakis and S.~S.~Pufu,
  ``Stability of strings binding heavy-quark mesons,''
  JHEP {\bf 0704} (2007) 079
  [arXiv:hep-th/0609137].
\bibitem{Avramis:2006nv}
  S.~D.~Avramis, K.~Sfetsos and K.~Siampos,
  ``Stability of strings dual to flux tubes between static quarks in N=4 SYM,''
  Nucl.\ Phys.\  B {\bf 769} (2007) 44
  [arXiv:hep-th/0612139].
\bibitem{Avramis:2007mv}
  S.~D.~Avramis, K.~Sfetsos and K.~Siampos,
  ``Stability of string configurations dual to quarkonium states in AdS/CFT,''
  Nucl.\ Phys.\  B {\bf 793} (2008) 1
  [arXiv:0706.2655 [hep-th]].
\bibitem{Sfetsos:2008yr}
  K.~Sfetsos and K.~Siampos,
  ``Stability issues with baryons in AdS/CFT,''
  JHEP {\bf 0808} (2008) 071
  [arXiv:0807.0236 [hep-th]].

\bibitem{working}
 M. Ali-Akbari, K. Bitaghsir,
 "Work in progress"

\bibitem{Athanasiou:2008pz}
  C.~Athanasiou, H.~Liu and K.~Rajagopal,
  ``Velocity Dependence of Baryon Screening in a Hot Strongly Coupled Plasma,''
  JHEP {\bf 0805} (2008) 083
  [arXiv:0801.1117 [hep-th]].
\bibitem{baryon}
  M.~Li, Y.~Zhou and P.~Pu,
  ``High spin baryon in hot strongly coupled plasma,''
  JHEP {\bf 0810}, 010 (2008)
  [arXiv:0805.1611 [hep-th]];
\bibitem{Sin:2009dk}
  S.~J.~Sin and Y.~Zhou,
  ``Holographic melting of Heavy Baryons in Plasma with Gluon Condensation,''
  arXiv:0904.4249 [hep-th]. C.~Krishnan,
  ``Baryon Dissociation in a Strongly Coupled Plasma,''
  JHEP {\bf 0812} (2008) 019
  [arXiv:0809.5143 [hep-th]].

\bibitem{Fadafan}
  K.~B.~Fadafan,
  ``Medium effect and finite 't Hooft coupling correction on drag force and Jet
  Quenching Parameter,''
  arXiv:0809.1336 [hep-th].
  K.~B.~Fadafan,
  ``$R^2$ curvature-squared corrections on drag force,''
  JHEP {\bf 0812} (2008) 051
  [arXiv:0803.2777 [hep-th]].
  J.~F.~Vazquez-Poritz,
  ``Drag force at finite 't Hooft coupling from AdS/CFT,''
   [arXiv:0803.2890 [hep-th]];

\bibitem{correction}
 M.T. Grisaru and D. Zanon, Phys. Lett B177 (1996) 347;

\bibitem{Banks}
  T.~Banks and M.~B.~Green,
  ``Non-perturbative effects in AdS(5) x S**5 string theory and d = 4 SUSY
  Yang-Mills,''
  JHEP {\bf 9805}, 002 (1998)
  [arXiv:hep-th/9804170];

\bibitem{alpha2}
 J. Pawelczyk and S. Theisen,
 {\it AdS$_5\times S^5$ black hole metric at {\cal O}($\alpha^{\prime 3}$)},
  JHEP {\bf 9809} (1998) 010, [hep-th/9808126];

\bibitem{alpha1}
 S.S. Gubser, I.R. Klebanov and A.A. Tseytlin,
{ \it Coupling constant dependence in the thermodynamics of $N=4$
supersymmetric Yang-Mills theory}
 Nucl. Phys. {\bf B534} (1998) 202, [hep-th/9805156];

\bibitem{Zwiebach}
  B.~Zwiebach,
  ``Curvature Squared Terms And String Theories,''
  Phys.\ Lett.\  B {\bf 156}, 315 (1985).

\bibitem{GB}
  R.~G.~Cai,
  ``Gauss-Bonnet black holes in AdS spaces,''
  Phys.\ Rev.\  D {\bf 65}, 084014 (2002)
  [arXiv:hep-th/0109133];

\bibitem{Kats:2007mq}
  Y.~Kats and P.~Petrov,
  ``Effect of curvature squared corrections in AdS on the viscosity of the dual
  gauge theory,''
  arXiv:0712.0743 [hep-th].
\bibitem{Buchel:2008vz}
  A.~Buchel, R.~C.~Myers and A.~Sinha,
  ``Beyond eta/s = 1/4pi,''
  JHEP {\bf 0903} (2009) 084
  [arXiv:0812.2521 [hep-th]].

\bibitem{Brigante:2008gz}
  M.~Brigante, H.~Liu, R.~C.~Myers, S.~Shenker and S.~Yaida,
  ``The Viscosity Bound and Causality Violation,''
  Phys.\ Rev.\ Lett.\  {\bf 100} (2008) 191601
  [arXiv:0802.3318 [hep-th]].
  X.~H.~Ge and S.~J.~Sin,
  ``Shear viscosity, instability and the upper bound of the Gauss-Bonnet
  JHEP {\bf 0905} (2009) 051
  [arXiv:0903.2527 [hep-th]].
\end{thebibliography}
\end{document}